**Chris Tofallis**[*]

# OBJECTIVE WEIGHTS FOR SCORING: THE AUTOMATIC DEMOCRATIC METHOD

## Abstract

When comparing performance (of products, services, entities, etc.), multiple attributes are involved. This paper deals with a way of weighting these attributes when one is seeking an overall score. It presents an objective approach to generating the weights in a scoring formula which avoids personal judgement. The first step is to find the maximum possible score for each assessed entity. These upper bound scores are found using Data Envelopment Analysis. In the second step the weights in the scoring formula are found by regressing the unique DEA scores on the attribute data. Reasons for using least squares and avoiding other distance measures are given. The method is tested on data where the true scores and weights are known. The method enables the construction of an objective scoring formula which has been generated from the data arising from all assessed entities and is, in that sense, democratic.

## 1 Introduction: Why have a formula for performance or efficiency?

We are often interested in comparing the performance or efficiency of entities, be they consumer products, services, people, organisations, etc. We naturally prefer results from techniques which are readily comprehensible because this provides greater confidence. Furthermore, this allows us to communicate the results more easily and to persuade others. Having a simple formula is therefore helpful. It also provides transparency: it clearly shows how the various factors are weighted and combined. The challenge that we shall investigate is deciding what these weights should be.

---

[*] University of Hertfordshire: Hatfield, GB, e-mail: ormsresearch@gmail.com, ORCID: 0000-0001-6150-0218.

The factors or attributes associated with the entities are often combined to produce a multi-dimensional or composite index and the weights chosen for these criteria are clearly crucial to the results that follow. A startling illustration of this is given by Decanq and Lugo (2013, p. 16) describing the work of Becker et al. (1987):

> "The authors studied the quality of life in 329 metropolitan areas of the U.S. by ordering them according to standard variables such as quality of climate, health, security, and economical performance. The authors find that, depending on the weighting scheme chosen, there were 134 cities that could be ranked first, and 150 cities that could be ranked last. Moreover, there were 59 cities that could be rated either first or last, using the same data, but by selecting alternative weighting schemes".

Data envelopment analysis (DEA) provides a way of deriving *separate* attribute weights for each entity so as to maximize its score, and thereby overcomes the issue of subjectivity. An entity assessed with a DEA score would be pleased with its allocated weights since no other weights could provide a higher score. However, some would argue that this removes a degree of comparability because each entity has its own weights and so comparison is not on the same basis. Hence, there is a perceived need, at least in some quarters, for a common set of weights that can be applied to all the entities being compared (Liu and Peng, 2008; Ramon, Ruiz and Sirvent, 2012; Kritikos, 2017). Liu and Lu (2010, p. 453) give an example of such a situation: "DEA usually provides a group of performance leaders that can be used as benchmarks for those who are outside the leading group. The leaders are of equal significance under the original DEA methodology. Some applications, however, expect unambiguous, preferably ranked, performance leaders. For example, when applying DEA to compare the performance of R&D (research and development) organizations, one prefers a ranked list in order to correctly reward the R&D organizations and more importantly to allocate precious resources to organizations with better performance". Ruiz and Sirvent (2016, p. 8) point out that "The DMUs [decision making units] involved in production processes often experience similar circumstances, so benchmarking analyses in those situations should identify common referents and establish common best practices (...). In particular, this means that input and output weights should be common to all units in the evaluations, in contrast to DEA".

By having common weights one can produce rankings based on the score from the formula. The media are often keen to present results from rankings, and so this is one way of gaining public attention, the attention of policy makers, and decision makers. Rankings are sometimes based on aggregating expert opinions, but these may be heavily influenced by historical reputations which may be less relevant today. In this regard Rosenthal and Weiss (2017, p. 136) look at academic journals in the field of business and notes that "While we believe that subjective comparisons and rankings are very important (…) we also believe that subjective rankings represent opinions that are often slow to change".

In magazines and newspapers, journalists will discuss shifting rank positions and comment on possible causes. Examples include the annual Human Development Index ranking of countries produced by the United Nations Development Programme, and the various national and international 'league tables' of universities which exist around the world. Despite being roundly criticised by academics, the latter attract a great deal of interest because of the power that they wield: potential students are influenced by them when selecting where to study, and academics seeking a position may also be guided by them. For their part, university executives scour these tables to see if they can trumpet recent successes when their position rises, or try to see which factors they need to improve if it falls. Another example is the Multidimensional Poverty Index, which is produced by the Oxford Poverty and Human Development Initiative (OPHI). They point out that a dashboard showing multiple measures does not reflect the multiple deprivations that some people face at the same time, whereas a single index can combine such information.

Other reasons for using a common set of weights relate to issues regarding the direct use of DEA. Perhaps the most serious issue from a managerial perspective is attaching zero (or epsilon) weights to criteria, thereby effectively hiding that factor from the assessment of that entity. Hence it is not unusual to have many DMUs with a score of 100%. It can be proved that if a unit has (uniquely) the largest value of a particular output then it is automatically efficient in some DEA models (e.g. the BCC or Additive DEA models) – and this is irrespective of how large (and possibly wasteful) its consumption of inputs! (Ali, 1993; Jahanshahloo et al., 2005). Similarly, if a unit has (uniquely) the smallest value of an input then it is automatically 100% efficient in some DEA models – irrespective of how low its output levels are! So, if there are $m$ inputs and $s$ outputs, up to $m + s$ units could be efficient in this way, which is an unfortunate artefact of such models.

Roll and Golany (1993, p. 99) point out that "it is usually deemed inappropriate to accord widely differing weights to the same factor, when assessing different DMUs". Indeed, it is this weight flexibility that leads to poor discrimination between entities following a DEA analysis, with many achieving a 100% score. This is especially true when the number of DMUs being compared is small, and/or the number of factors (inputs, outputs, performance measures) is large.

A graphical representation may assist in understanding how DEA may accord widely differing weights. In Figure 1 all units are assumed to have the same input value; the four upper DMUs all produce more than 20 units of the output on the vertical axis, whereas the fifth DMU produces one unit only. Nevertheless, it is rated 100% under DEA due to the fact that it has the highest level of the output on the horizontal axis. Such units are sometimes referred to as mavericks. Moreover, any DMUs just behind this one will have scores close to 100%. The relative weights accorded to the two outputs will clearly differ greatly as indicated by the slope of the line-segment on the right compared to the slopes of the other sections of the frontier.

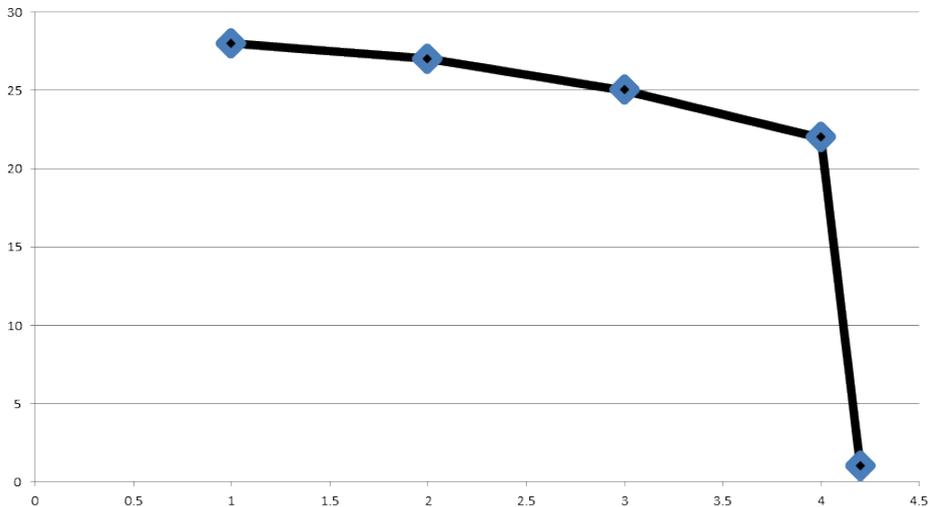

Figure 1: Frontier units in output space

## 2   Common weights and DEA

Perhaps the first paper to propose a common set of weights (CSW) in the DEA context is Cook, Roll and Kazakov (1990), which mentions it as a direction for further analysis. This was followed by Roll, Cook and Golany (1991, p. 6), who motivate their work by pointing out that using a single set of weights "(…) is the usual approach in all engineering, and most economic, efficiency analyses. In these cases it is assumed that all important factors affecting performance are included in the measurement system, and there is no need (nor wish) to allow for additional, individual, circumstances". However, they point out that in reality some factors will not have been included in the analysis, and that there may also be differences in goals and missions. Roll, Cook and Golany (1991, p. 6) then propose an interesting interpretation:

> "A possible meaning of efficiencies computed with a CSW, in the context of DEA, is that such values represent the part of a DMU's performance which can be explained when assuming uniformity of circumstances. The difference between the efficiency measured with an »individual« set of weights and that obtained with a CSW may indicate the effects of special circumstances under which a DMU operates".

Among the approaches that they suggest is to run a conventional DEA and then take the average weight attached to each factor across all the DMUs, or to take some average of the highest and lowest weight. This is a move away from DEA as 'self-appraisal', and towards 'peer appraisal'. This is the philosophy behind 'cross-efficiency'; here the evaluation of an entity is some 'average' of the evaluations arising from applying the optimal weights of all the entities to the one being assessed. The problem with this is that DEA weights are not unique – one can obtain the same DEA score using different weights. To quote Thompson, Dharmapala and Thrall (1993, p. 383) "Zero values in optimal solutions for the primal or dual LP programs are indicators of degeneracy and multiple optimal solutions. Some analysts misleadingly use »the« instead of »a« or »an« for a DEA solution; and accordingly, they may fail to recognize the existence of multiple optimal solutions". The non-uniqueness of DEA weights is a long-running issue with the cross-efficiency method (Wu, Sun and Liang, 2020), and has led to a plethora of sophisticated devices for dealing with them. Indeed, Table 1 in Contreras, Lozano and Hinojosa (2021) lists 44 cross-efficiency approaches. For the sake of simplicity we choose to avoid this route.

Roll and Golany (1993) presented further ideas for determining common weights. One suggestion was to choose the CSW which maximizes the average score of the DMUs. Another was to determine the CSW which maximized the number of efficient DMUs. We shall not attempt here to review the extensive literature on CSW, as this has been carried out by Aldamak and Zolfaghari (2017), which contains 113 references − there is also a review of ranking models in DEA by Lotfi et al. (2013) with 104 references, see also Adler, Friedman and Sinuany-Stern (2002), and Moghaddas and Vaez-Ghasemi, 2017. Nevertheless, we now mention those papers which bear some similarity to the work we shall present.

Kao and Hung (2005) treated the DEA scores as the ideal solution to be used in a compromise programming formulation to identify the common weights: this involves finding a solution which is as close as possible to the ideal point. The three most common distance measures were used: city-block (or Manhattan), Euclidean, and Chebyshev ($L_1$, $L_2$, and $L_\infty$ metrics). They recommended investigating all three of these and then to make a subjective judgement. Surprisingly, they reported that replacing all the DEA scores by 100% led to the same CSW for the $L_1$ metric. We now explain why this happens: The objective in the $L_1$ metric is to find a common set of weights ($u,v$) so as to minimize the sum of absolute differences between the DEA scores ($E_i^*$) and those from the CSW formula:

$$\text{Min } \Sigma |E_i^* - E_i(u,v)| \tag{1}$$

The DEA scores are optimal and therefore should not be exceeded by using any other set of weights, so $E_i^* \geq E_i(u,v)$, which implies that the objective simplifies to:

$$\text{Min } \Sigma[E_i^* - E_i(u,v)] \text{ i.e. Min } [\text{constant} - \Sigma E_i(u,v)] \tag{2}$$

Thus the actual DEA scores are removed from the problem and do not affect the resulting common set of weights. This is rather disheartening and leads us to reject this $L_1$ approach. (The objective function also simplifies to maximising the sum of scores.)

Zohrehbandian, Makui and Alinezhad (2010) adapted Kao and Hung's approach to avoid nonlinearities by restricting attention to the $L_1$ and $L_\infty$ metrics, and by considering an additive DEA model. Pre-dating both of these papers is Despotis (2002), who combines the $L_1$ and $L_\infty$ distance measures using a weighting parameter which is left to the user to set. This was later applied to the Human Development Index (Despotis, 2005).

Cook and Zhu (2007) consider the view that one should minimize the deviation for the unit whose final score is furthest from its original DEA score. This is an $L_\infty$ or Chebyshev metric and so we have a minimax approach. This view could be supported on the grounds that this unit is the most disadvantaged. However, one needs to ask why there is a large deviation in score. It may be that this unit is a maverick as in Figure 1 and is using weights which differ considerably from the rest of the data set, or even had zero weights on a number of variables and so was able to hide the extent of its inefficiency; thus the more realistic (common) weights caused its score to decline the most. It therefore becomes highly questionable whether such a poor performing unit should be so influential in determining the weights for all other units. It also appears to be 'undemocratic' that a single unit should hold such power over the others. It is for these reasons that we do not adopt the $L_\infty$ metric.

## 3  The automatic democratic method

In this paper we propose applying least squares regression to the DEA scores ($E_i$) as the dependent variable. For reasons stated above we do not use the $L_1$ and $L_\infty$ fitting approaches. The underlying attribute data are used as the explanatory variables in a functional form that reflects the efficiency measure used in DEA when finding the optimal score for each assessed entity. We shall allow for both input (more is worse) and output (more is better) variables. One way of measuring performance efficiency is the ratio: sum of weighted outputs ($y$) to sum of weighted inputs ($x$), which is the original CCR model for DEA (Charnes, Cooper and Rhodes, 1978). We emphasise that any other functional form (additive, multiplicative etc.) can be used as the performance metric – our approach is not restricted in this way. For our ratio form the following least squares objective function would apply:

$$\text{Minimize} \sum_{i=1}^{n} \left[ \frac{\sum_{r=1}^{s} u_r y_{ri}}{\sum_{j=1}^{m} v_j x_{ji}} - E_i \right]^2 \qquad (3)$$

where $u_r$ is the coefficient for output $r$ and $v_j$ is the coefficient for input $j$, these are the common weights to be determined by the regression, and the $E_i$ are the numerical scores previously obtained from DEA for each unit $i$. The idea is to produce a handy compact formula for the scores. This can be described as an Automatic Democratic Approach since each entity is *initially* represented by its own optimal score, with individual weights emphasising its best aspects, with the *final* set of weights and scores being deduced objectively from these initial scores by regression.

Given that DEA scores are designed to show each entity in the best possible light, and are therefore optimistic, we feel that such scores ought to be treated as upper bounds, and not exceeded. We thus take measures to deal with this issue. Two ways will be considered:

1. Use constrained regression with one-sided residuals. The constraints which force all residuals to have the same sign guarantee that DEA scores (as the dependent variable) will not be exceeded.
2. Use ordinary regression followed by an adjustment which ensures the formula score does not exceed the DEA score. This can be done by subtracting a constant (equal to the largest over-estimate) from all scores; however, this runs the risk of making very low scores become negative. A better approach is to rescale i.e. divide the formula scores by a factor which ensures our requirement. The factor will be the largest instance of the ratio (DEA score)/(Formula score). This approach also has the benefit of retaining proportionality between the formula scores.

It should be noted that the regression we are discussing here differs from the commonly used second stage regressions where DEA scores are related to other variables (e.g. environmental, contextual, or otherwise explanatory) not used in estimating the DEA scores. Such models sometimes employ a censored normal Tobit specification to deal with the fact that there is a concentration of observations with a 100% score. Such points are an artefact of the DEA method of identifying a frontier: attaching a score of 100% to some observations should not make us think there is anything absolute or fixed about them, since DEA is always an indication of 'relative' performance.

We stress that we are using regression for descriptive purposes – to provide a formula which approximates a set of scores. There is no assumption regarding the underlying probability distribution of the scores, and no statistical inference is being made.

### 3.1 Illustrative application

It has been claimed that "Data Envelopment Analysis (DEA) is the most commonly used approach for evaluating healthcare efficiency" (Hollingsworth, 2008; Gajewski et al., 2009; Matawie and Assaf, 2010), and we shall use an example from this field to illustrate our method. The data (Cooper, Seiford and Tone, 2006, p. 169) cover 14 general hospitals and have two inputs: nurses and doctors. The two outputs are inpatient and outpatient numbers (see Table 1). Applying DEA using the CCR model we obtain the scores shown in Table 2. Also shown are those weights which are zero or extremely small. What is disturbing is that half of the hospitals have a zero weight on the number of outpatients! Also, four hospitals place zero weight on doctors. These variables are effectively being ignored in assessing those hospitals. Commenting on this, Cooper, Seiford and Tone (2006, p. 169) state: "This means that the inefficient hospitals are very likely to have a surplus in doctors and a shortage of outpatients". There are even three hospitals which place zero weights on both doctors and outpatients, and so are being assessed purely on the ratio of inpatients per nurse, with the other variables ignored. This highlights how DEA, when used on its own, can be problematic when evaluating performance.

Table 1: Data on 14 general hospitals

| Hospital | Doctors | Nurses | Outpatients | Inpatients |
|---|---|---|---|---|
| A | 3008 | 20 980 | 97 775 | 101 225 |
| B | 3985 | 25 643 | 135 871 | 130 580 |
| C | 4324 | 26 978 | 133 655 | 168 473 |
| D | 3534 | 25 361 | 46 243 | 100 407 |
| E | 8836 | 40 796 | 176 661 | 215 616 |
| F | 5376 | 37 562 | 182 576 | 217 615 |
| G | 4982 | 33 088 | 98 880 | 167 278 |
| H | 4775 | 39 122 | 136 701 | 193 393 |
| I | 8046 | 42 958 | 225 138 | 256 575 |
| J | 8554 | 48 955 | 257 370 | 312 877 |
| K | 6147 | 45 514 | 165 274 | 227 099 |
| L | 8366 | 55 140 | 203 989 | 321 623 |
| M | 13 479 | 68 037 | 174 270 | 341 743 |
| N | 21 808 | 78 302 | 322 990 | 487 539 |

Table 2: DEA scores based on the CCR method applied to 14 hospitals

| Hospital | DEA score | Doctor weight | Nurse weight | Outpatient weight | Inpatient weight |
|---|---|---|---|---|---|
| A | 0.955 | | 0 | | |
| B | 1 | | | | |
| C | 1 | | | | |
| D | 0.702 | | | 0 | |
| E | 0.827 | 0 | | 0 | |
| F | 1 | | | | |
| G | 0.844 | | | 0 | |
| H | 1 | | 4.6 E-08 | 1.1 E-08 | |
| I | 0.995 | 0 | | | |
| J | 1 | | | | |
| K | 0.913 | | 7.4 E-08 | 0 | |
| L | 0.969 | | | 0 | |
| M | 0.786 | 0 | | 0 | |
| N | 0.974 | 0 | | 0 | |

The table also shows those weights on inputs and outputs that turn out to be zero or extremely small.

We now apply the proposed automatic democratic method. We regress the DEA scores as the dependent variable on the underlying input and output data, as explanatory variables. The ordinary least squares (OLS) regression model is:

DEA score = [$u_1$ Outpatients + $u_2$ Inpatients]/[$v_1$ Doctors + $v_2$ Nurses] + residual  (4)

This is a nonlinear regression in the coefficients (or weights $u$, $v$), but is easily solved using standard statistical software such as SPSS.

The resulting formula is:

E = [Outpatients + 4.94 Inpatients] / [52.18 Doctors + 24.56 Nurses]   (5)

where we have set the coefficient for Outpatients to be 1 for ease of interpretation; there is no loss of generality as the scores remain unchanged when all coefficients are divided by the same number.

First we are pleased to observe that there are no zero weights. Inspecting the output weights shows that an inpatient has a greater weight than an outpatient, which is as expected. Likewise, for the inputs, a doctor has a greater weight than a nurse; again this is what we would expect.

Table 3: Comparing ranks using DEA, OLS (Ordinary Least Squares), and a constrained regression formula

| Hospital | A | **_B_** | **_C_** | D | E | **_F_** | G | **_H_** | I | **_J_** | K | L | M | N |
|---|---|---|---|---|---|---|---|---|---|---|---|---|---|---|
| OLS rank | 10 | 6 | 2 | 14 | 12 | 3 | 11 | 7 | 4 | _1_ | 8 | 5 | 13 | 9 |
| DEA rank | 9 | _1_ | _1_ | 14 | 12 | _1_ | 11 | _1_ | 6 | _1_ | 10 | 8 | 13 | 7 |
| Constrained regression | 11 | 9 | 2 | 14 | 12 | 3 | 10 | 6 | 5 | _1_ | 7 | 4 | 13 | 8 |

DEA-efficient hospitals are underlined.

In Table 3 we compare the ranks based on DEA with those arising from the formula. Most of the hospitals have the same or similar rank, which is reassuring. However, we need to look at those cases where there is a marked difference and see if this makes sense. The largest change occurs with hospital H which was 100% efficient under DEA but is now ranked 7$^{th}$ under OLS. To explain this decline let us compare it with C which has rank 2. Note that the CCR model assumes constant returns to scale. H treats 15% more inpatients and just 2% more outpatients, but H is outperformed by C because H uses 45% more nurses to achieve this. Inspection of the weights in Table 2 shows that H has placed a negligible weight on nurses, thereby downplaying this high resource usage.

Next consider hospital B which was DEA efficient, but now has rank 6. Let us try and understand this by comparing it with C (ranked second). C has 5% more nurses and 8.5% more doctors than B, and so we would expect its processing of inpatients to be greater by this order, but it is actually processing 29% more inpatients than B. This disproportionate difference helps explain why C now outperforms B.

Next we turn to the (unconstrained) OLS scores provided by the formula. Given that it is an approximation based on regression, it is inevitable that some scores will be above or below their DEA scores, as can be seen in Table 4. As DEA scores are viewed as an optimistic upper bound, there is no requirement with formula scores which are below these. However, something needs to be done if the formula provides a score higher than the corresponding DEA value. This is fixed by a simple rescaling. For each hospital we calculate the ratio (DEA score)/(Formula score), then find the maximum ratio. We then divide all the formula scores by this maximum ratio. This ensures that DEA scores are never exceeded whilst also maintaining proportionality: all scores remain the same in relative terms, so that ratios of scores are unchanged.

Table 4: Comparison of DEA and formula scores

| Hospital | A | B | C | D | E | F | G | H | I | J | K | L | M | N |
|---|---|---|---|---|---|---|---|---|---|---|---|---|---|---|
| DEA score | 0.96 | 1.00 | 1.00 | 0.70 | 0.83 | 1.00 | 0.84 | 1.00 | 1.00 | 1.00 | 0.91 | 0.97 | 0.79 | 0.97 |
| OLS Formula score | 0.89 | 0.93 | 1.09 | 0.67 | 0.85 | 1.05 | 0.86 | 0.90 | 1.01 | 1.09 | 0.89 | 1.00 | 0.78 | 0.89 |
| OLS Formula score (rescaled) | 0.81 | 0.85 | 0.99 | 0.61 | 0.78 | 0.96 | 0.79 | 0.83 | 0.93 | 1.00 | 0.82 | 0.92 | 0.72 | 0.82 |
| Constrained regression score | 0.80 | 0.83 | 1.00 | 0.65 | 0.77 | 0.96 | 0.82 | 0.84 | 0.92 | 1.00 | 0.83 | 0.94 | 0.75 | 0.83 |

Our alternative approach for ensuring that formula scores do not exceed DEA scores is to use regression with constraints to ensure the residuals are one-sided, i.e. all have the same sign. The resulting formula for this is:

$$E = [\text{Outpatients} + 60 \text{ Inpatients}] / [628 \text{ Doctors} + 279 \text{ Nurses}] \qquad (6)$$

We note the relative weights for inpatients and outpatients are now 60 to 1, which is much higher than before. For doctors and nurses the relative weights are only slightly altered. From the bottom two rows of Table 3 we see that all but one of the ranks are the same or differ by one from the earlier formula; and the bottom two rows of Table 4 indicate similar scores using the two formulae. It is interesting to compare the constrained regression results with those from the earlier formula by looking at deviations from the DEA scores. Note that all the deviations have the same sign. The largest residual was 0.175 in both cases. More interesting was that the mean deviation using constrained regression was 0.0725, whereas it was higher, at 0.082, using the OLS formula with rescaling. To understand why this should be the case provides a useful insight which enables us to choose between these approaches. We expect the unconstrained OLS 'predictions' to be scattered above and below the DEA values in a roughly symmetric random manner. However, the effect of the subsequent adjustment by rescaling will extend the deviation associated with those points which were already under-predicted, i.e. they will be dragged down even further. This effect will, in general, cause constrained regression to provide a closer fit than OLS followed by rescaling because it takes into account the one-sided condition at the same time as the optimal fitting process.

Another way to improve the fit is to include a constant or intercept in the model formula. For the constrained regression this causes the mean deviation to fall to 0.039, which is quite a drop from 0.0725. The largest residual was 0.169. For OLS with rescaling the mean deviation was 0.066 and the largest deviation was 0.157.

We point out that whilst we have assumed a formula which is similar in form to that used to assess the units in DEA, this does not have to be the case. One could achieve an even closer fit using a more flexible form. A ratio of linear forms assumes constant rates of marginal substitution, and no curvature in the frontier. But a ratio of quadratics (or other nonlinear forms) would allow for non-constant substitution rates and curved frontiers.

### 3.2  Testing the method where the true weights and scores are known

In general, we would not know the 'true' weights or 'true' scores. However, one can set up a situation where there is an assumed pre-specified relationship and then generate data. An example of this is given by Bowlin et al. (1985), where they considered hospitals being assessed on three outputs: the number of regular patients (RP), number of severe patients (SP), and number of teaching units (TU; student nurses, interns, etc.), while the input was the total cost (in \$'000), which was assumed to be related to the outputs according to:

$$\text{Efficient Cost} = 0.5\,TU + 0.13368\,RP + 0.17474\,SP \qquad (7)$$

assuming the hospital is operating efficiently. They generated data for seven efficient hospitals and eight hospitals operating inefficiently (Table 5). The true score or efficiency can be found from the ratio Efficient Cost/Actual Cost, i.e.:

$$\text{True Score} = [0.5\,TU + 0.13368\,RP + 0.17474\,SP]/\,\text{Cost} \qquad (8)$$

Table 5: Test data on 15 hospitals with true scores compared with estimated scores

| HOSPITAL | Teaching units | Regular Patients | Severe Patients | Cost $'000 | True Score | Automatic-Democratic Score |
|---|---|---|---|---|---|---|
| H1 | 50 | 3000 | 2000 | 775.5 | 1 | 0.995 |
| H2 | 50 | 2000 | 3000 | 816.6 | 1 | 0.992 |
| H3 | 100 | 2000 | 3000 | 841.6 | 1 | 0.992 |
| H4 | 100 | 3000 | 2000 | 800.5 | 1 | 0.995 |
| H5 | 50 | 3000 | 3000 | 950.3 | 1 | 0.994 |
| H6 | 100 | 2000 | 5000 | 1191.05 | 1 | 0.990 |
| H7 | 50 | 10000 | 2000 | 1711.3 | 1 | 1.000 |
| H8 | 100 | 3000 | 2000 | 884.75 | 0.91 | 0.900 |
| H9 | 50 | 2000 | 3000 | 841.6 | 0.97 | 0.963 |
| H10 | 100 | 10000 | 2000 | 2036.3 | 0.85 | 0.852 |
| H11 | 50 | 5000 | 3000 | 1362.6 | 0.89 | 0.890 |
| H12 | 100 | 3000 | 3000 | 1070 | 0.91 | 0.905 |
| H13 | 50 | 4000 | 5000 | 1491.1 | 0.96 | 0.955 |
| H14 | 100 | 3000 | 3000 | 1070 | 0.91 | 0.905 |
| H15 | 50 | 3000 | 2000 | 898.7 | 0.86 | 0.859 |

Source: Bowlin et al. (1985).

DEA was applied to the data in Table 5 to maximise the score ($u_1$TU + $u_2$RP + $u_3$SP) / Cost, for each hospital. We then applied constrained least squares regression to obtain common weights. The resulting scoring formula was found to be:

Automatic Democratic Score = [0.488 TU + 0.13420 RP + 0.17243 SP] / Cost.    (9)

We immediately notice that the weights are very close to those in the True Score formula above. The associated scores are shown in the last column of Table 5 and can be seen to be extremely close to the true values.

Thanassoulis (1993) applied regression directly to the above data using the cost as the dependent variable, and found:

$$\text{Cost} = 1.1054\ \text{TU} + 0.14811\ \text{RP} + 0.16198\ \text{SP} \qquad (10)$$

These coefficients are further away from those in the underlying equation (7). This is to be expected because regression, when used alone, will model *average* behaviour, not relative performance. It is also worth noting that applying regression directly to the data is restricted to situations where there is a single dependent variable, and so cannot be applied when there are both multiple input and multiple output variables. By contrast, in our two-step approach we have used DEA to assess performance (a single variable score), and then used regression to model this performance (Tofallis, 2013).

## 4  Conclusion

Our aim has been to provide a straightforward method for producing a formula for performance when there are no agreed weights on the underlying criteria. Initially, scores are obtained using DEA (which, by itself, does not generate a scoring formula). These are scores that each entity would be pleased with, since DEA uses the weights which maximizes its score. Consequently, DEA scores are sometimes felt to be unrealistically high, so we view these as optimistic, and treat them as upper bounds when constructing the scoring formula. The DEA scores are regressed on the underlying attribute data in order to objectively generate a common set of weights (the regression coefficients). We thus have an 'automatic democratic' approach in that data from all entities being assessed play a part in obtaining the weights.

Whilst the DEA literature contains much work on obtaining common weights, the method presented here is simpler and more direct. For example, much research has been done using cross-efficiency (see Wu, Sun and Liang, 2020, for a review), which involves aggregating scores using weights from self and peer perspectives; however, the fact that such weights are not unique requires additional methodological complications, such as secondary goals, in order to deal with this issue. By focusing on DEA scores (which are unique) rather than the associated weights (which are not) we avoid complications arising in many other approaches. Furthermore, using least squares regression brings the benefits of a well-known and widely used technique to provide a way of generating a compact formula for performance. Use of a formula also helps overcome the lack of discrimination associated with DEA, where many units have unreasonably high or even 100% efficiency as a result of DEA's extreme weight flexibility.

Our view is that the DEA scores are optimistic upper bounds, and so we have ensured this condition is upheld. Two ways were investigated for achieving this. We compared results using constrained regression with those from OLS followed by rescaling, and found that the former provides a closer fit. An explanation was provided for this effect: Constrained regression, by definition, fits as close as possible while maintaining the required condition; whereas making an adjustment after OLS does not carry the same 'best fit' property because the optimisation and constraints are not dealt with simultaneously. This insight allows us to reject the rescaling approach. We also considered the Manhattan and Chebyshev distance metrics as alternatives to least squares for regression and identified reasons for not adopting them.

The method we have presented has the attractive property of allowing a ranking of entities without requiring the imposition of arbitrary choices or restrictions on weights. Some of the DEA ranking literature is focused on achieving what is called a 'full ranking', that is one in which there are no ties. But it has to be realised that, if a formula is to be used, there will always be the possibility of two units with different attribute levels achieving an equivalent overall score.

The functional form of the scoring formula is not restricted by our approach. Although we have used a ratio of weighted outputs to weighted inputs as our illustrative example, it is of course possible to apply regression to scores generated by other models. For example, the BCC (Banker, Charnes, Cooper) model has an additional parameter to model variable returns to scale, and so the associated regression equation would provide a closer fit to the DEA scores. There are also additive and multiplicative DEA models which can be used for generating additive (linear) and multiplicative scoring formulae.

Finally, a valuable benefit of the proposed approach is the transparency of using a simple, and objectively constructed formula, for performance evaluation.